\documentclass[epsfig,12pt,onecolumn]{article}
\usepackage{amsfonts}
\usepackage{amssymb}
\usepackage{amsmath}
\usepackage{multicol}
\usepackage{graphicx}
\usepackage{float}
\usepackage{caption}

\input{tcilatex}
\makeatletter \@addtoreset{equation}{section}

\def \be{\begin{equation}}
\def \ee{\end{equation}}
\def \bea{\begin{eqnarray}}
\def \eea{\end{eqnarray}}

\newcommand{\nc}{\newcommand}
\nc{\al}{\alpha} \nc{\bib}{\bibitem} \nc{\la}{\lambda}
\nc{\C}{\mbox{\hspace{1.24mm}\rule{0.2mm}{2.5mm}\hspace{-2.7mm} C}}
\nc{\R}{\mbox{\hspace{.04mm}\rule{0.2mm}{2.8mm}\hspace{-1.5mm} R}}

\begin{document}

\title{\textbf{Topological densities\ in Einstein-scalar-Gauss-Bonnet}$\ $%
\textbf{gravity}}
\author{M. Bousder$^{1}$\thanks{%
mostafa.bousder@gmail.com} and Z Sakhi$^{2}$ \\
$^{1}${\small Faculty of Science, Mohammed V University in Rabat, Rabat,
Morocco}\\
$^{2}${\small Quantum Physics and Applications Team, LPMC, Faculty of
Science Ben M'sik,}\\
{\small Casablanca Hassan II University, Morocco}}
\maketitle

\begin{abstract}
The present work is devoted to studying the background dynamical evolution
of a scalar field in Einstein-Gauss-Bonnet gravity in maximally symmetric
space-time. This study is useful for giving meaning to the presence of two
Gauss-Bonnet vacua, instead of using the spherically symmetric bubbles of
the "true" vacuum expand in the "false" vacuum. The theory admits two
possible effective cosmological constants, which lead to two maximally
symmetric vacuum solutions. The first solution corresponds to the dynamics
of dark energy. When there is matter, the second solution describes dark
matter. In Einstein-Gauss-Bonnet gravity, we establish the expression of the
topological mass spectrum which depends on the golden ratio and its inverse.
In the Schwarzschild limit, the topological density corresponds to the
standard model radiation energy density. We find the mass loss rate which
gives the evolution of mass over time.
\end{abstract}

\section{ Introduction}

The observations of the rotation of galaxies and gravitational lenses
indicate the presence of a new mass, called dark matter (DM) hiding in
galaxies, which does not interact with radiation and matter, but can be
detected by its gravitational effect. The $\Lambda $CDM model is a
cosmological model, parametrized by a cosmological constant $\Lambda $\
associated with cold dark matter. It is often called the standard Big Bang
model because it is the simplest model that accounts for the properties of
the cosmos: the large-scale structure of the observable universe and the
distribution of galaxies, the abundance of light elements (hydrogen, helium
and lithium) and the expansion of the universe. $\Lambda $CDM model assumes
that general relativity theory correctly describes gravity on a cosmological
scale. However, the $\Lambda $CDM model presents several problems, such as
the cosmological constant, fine-tuning problem, and the problem of cosmic
coincidence \cite{0000}.\ Recent developments of non-trivial extension of
Lovelock theory, namely Einstein-Gauss-Bonnet (EGB) theory have been
proposed \cite{G1}, providing new insights into the 4-dimensional theory of
gravity \cite{DE}. Their idea is that before taking the limit $%
D\longrightarrow 4$, they multiplied the Gauss-Bonnet (GB) term by the
factor $1/(D-4)$. The divergent factor $1/(D-4)$ is canceled by the
vanishing GB contributions, which leads to a theory of gravity with only two
dynamical degrees of freedom, which is in contradiction with Lovelock
theorem \cite{R1} which describes the gravity at $D\geq 5$.\ However, it was
shown in several papers that the idea of the limit $D\longrightarrow 4$ is
not clearly defined, as well as the absence of proper action \cite%
{F1,F2,F3,F4}. It was explicitly confirmed by a direct product $D$%
-dimensional spacetime or by adding a counter term, before taking the limit $%
D\longrightarrow 4$, which can be seen as a class of Horndeski theory \cite%
{N0} but with $2+1$-dofs. Although the EGB gravity is currently debatable,
the spherically symmetric black hole solution is still meaningful and worthy
of study \cite{N1}. \newline
The 4D symmetrical static and spherical black hole solution in EGB gravity
were obtained \cite{N9}, also, solutions of static and spherically symmetric
compact stars \cite{N10,N11,N12}. Many researchers have studied the
mass-radius profile and the maximum mass in EGB gravity \cite{N13,N14}.
Consequently, it is possible to describe the matter inside compact objects
(COs) and the dynamical evolution of the matter at high density and the
behavior of violent events. The Einstein-scalar-Gauss-Bonnet (EsGB) gravity%
\textbf{\ }\cite{NN1}, are the class of classical scalar-tensor theories
that have second-order EOMs, as a special case the Horndeski gravity \cite%
{NN2}. The EGB gravity admits two maximally symmetric vacuum solutions as
the Einstein vacuum in $\alpha \rightarrow 0$, and the Gauss-Bonnet vacuum
in $\alpha \neq 0$ \cite{BB,B3}. Previous studies \cite{133} including the
effect of varying the cosmological constant, showed that the correspondence
between ordinary thermodynamic systems and black hole mechanics would be
completed to include a variable cosmological constant. The bubble nucleation
probability depends on the curvature coupling of the Higgs fields, which is
a renormalizable parameter of the Standard Model (SM) in curved spacetime
\cite{NN3}. In the effective field theory, the thermal and quantum
fluctuations to overcome the barrier are characterized by the decrease of
the vacuum. The bubble nucleation in thermal fluctuations can be described
in terms of Euclidean time coordinate by instantons \cite{NN4}. The action
of the Coleman-de Luccia instanton determines the rate of vacuum decay \cite%
{NN5}. The classic solutions for switching from a false vacuum to a true
vacuum are called instantons. We suppose that the Boltzmann constant $k_{B}$%
, the reduced Planck constant $\hbar $ and light speed $c,$ are such that: $%
k_{B}=\hbar =c=1.$\newline
We begin in section 2 with a discussion of\ the equation of motion for a
coupling between a scalar field and the model.\ the chameleon mechanism and
the scalaron mass are investigated in section 3. Section 4 is devoted to the
two branches of solutions for the effective cosmological constant in a
maximally symmetric vacuum.\ In Section 5, we study the scalar dark matter
in the exterior region of CO by the EsGB and the regions of validity of the
GB functional coupling. In Section 6, we explore the applications for the 4D
Einstein-Gauss-Bonnet black hole. Finally, the paper ends in Section 7 by
summarizing our main results.

\section{Einstein-scalar-Gauss-Bonnet gravity}

We start by the action of Einstein-scalar-Gauss-Bonnet (EsGB) gravity \cite%
{D1,D2} in $4$-dimensions as

\begin{equation}
\mathcal{S}=\int d^{4}x\sqrt{-g}\left( \frac{M_{p}^{2}}{2}R+f\left( \phi
\right) \mathcal{G}+\mathcal{L}_{DM}\right) +\mathcal{S}_{m},  \label{EG1}
\end{equation}%
where $S_{m}$ is the matter action, $M_{p}=\frac{c^{4}}{8\pi G}\approx
2\times 10^{18}\left[ GeV\right] $, $R$ is the Ricci scalar, $f\left( \phi
\right) $ is a Gauss-Bonnet coupling function, which is ultraviolet (UV)
corrections to Einstein theory. We define the Gauss-Bonnet invariant \cite%
{MBOU} as
\begin{equation}
\mathcal{G}=R^{2}-4R_{\mu \nu }R^{\mu \nu }+R_{\mu \nu \rho \sigma }R^{\mu
\nu \rho \sigma }.  \label{EG2}
\end{equation}%
The variation with respect to the field $\phi $ gives us the equation of
motion for the scalaron field%
\begin{equation}
\square \phi -\partial _{\phi }V\left( \phi \right) +\mathcal{G}\partial
_{\phi }f\left( \phi \right) =0.  \label{EG4}
\end{equation}%
The variation of the action over the metric $g_{\mu \nu }$ simplified by the
Bianchi identity gives the equations of motion in \cite{D1}. In the Jordan
frame,\textbf{\ }the scalar DM Lagrangian reads \cite{QW1}%
\begin{equation}
\mathcal{L}_{DM}=-\frac{1}{2}g^{\mu \nu }\partial _{\mu }\phi \partial _{\nu
}\phi -V\left( \phi \right)  \label{EG5}
\end{equation}%
the kinetic term is invariant under the transformation $\phi \rightarrow
-\phi $ of the $Z_{2}$ symmetry. The metric of a spatially flat homogeneous
and isotropic universe in FLRW model is given by:%
\begin{equation}
ds^{2}=-dt^{2}+a^{2}(t)\sum_{i=1}^{3}\left( dx^{i}\right) ^{2},  \label{EG6}
\end{equation}%
where $a(t)$ is a dimensionless scale factor, from which we define the Ricci
scalar $R$ and the GB invariant $\mathcal{G}$ in FLRW geometry as%
\begin{equation}
R=6\left( 2H^{2}+\dot{H}\right) \text{ \ \ \ \ }\mathcal{G}=24H^{2}\left(
\dot{H}+H^{2}\right) .  \label{RR}
\end{equation}%
We start by considering $\phi =\phi \left( t\right) $. where $\dot{\phi}%
=\partial _{t}\phi $, $f^{\prime }\left( \phi \right) =\partial _{\phi
}f\left( \phi \right) $ and $H=\dot{a}/a$ is the Hubble parameter. Eq.(\ref%
{EG4}) can be written as the Klein Gordon equation in a simple form%
\begin{equation}
\ddot{\phi}+3H\left[ \dot{\phi}-8f^{\prime }\left( \phi \right) H\left( \dot{%
H}+H^{2}\right) \right] +V^{\prime }\left( \phi \right) =0.  \label{EG8}
\end{equation}

\section{Scalaron mass in EsGB gravity}

Recently, there has been a renewed interest in the relationship between dark
matter and the scalaron mass (i.e. the mass of fields $\phi $) \cite{40}. We
will study later the scalaron mass $m_{\phi }$ which will describe the dark
matter. To describe the mass of any scalar field, we use the Klein-Gordon
equation $\square \phi =\partial _{\phi }V_{eff}(\phi )$. Note that Eq.(\ref%
{EG4}) gives the same form as in the last equation. Since $\mathcal{G}$ in
Eq.(\ref{EG4}), does not depend on $\phi $, in this case, the effective
potential is evaluated as follows%
\begin{equation}
V_{eff}(\phi )=V\left( \phi \right) -f\left( \phi \right) \mathcal{G}.
\label{acd}
\end{equation}%
We notice that the effective potential of the scalaron includes the
Gauss-Bonnet coupling and the Gauss-Bonnet invariant. In other words, the
Gauss-Bonnet term affects the potential structure of the scalaron; thus, the
scalaron mass depends on the matter's contribution. The particles of the
field $\phi $ come from the fluctuation around the minimum of the effective
potential $V_{eff}(\phi )$. The mass of small fluctuations around $\phi
_{\min }$ Eq.(\ref{acc}) give the scalaron mass \cite{MBOU2} is determined as%
\begin{equation}
m_{\phi }^{2}=\frac{\partial ^{2}}{\partial \phi ^{2}}V_{eff}(\phi _{\min }),
\label{d}
\end{equation}%
where $V_{eff}(\phi _{\min })$\ as a minimum value of the scalaron effective
potential $V_{eff}$. Also we have $\mathcal{G}=\frac{\partial _{\phi
}^{2}V\left( \phi \right) }{\partial _{\phi }^{2}f\left( \phi \right) }-%
\frac{\partial _{\phi }^{2}V_{eff}(\phi )}{\partial _{\phi }^{2}f\left( \phi
\right) }$. To make progress, let us express the $\mathcal{G}$ in a more
convenient form%
\begin{equation}
\mathcal{G}=\frac{\partial _{\phi }^{2}V\left( \phi _{\min }\right) }{%
\partial _{\phi }^{2}f\left( \phi _{\min }\right) }-\frac{m_{\phi }^{2}}{%
\partial _{\phi }^{2}f\left( \phi _{\min }\right) }.  \label{g1}
\end{equation}%
For $f\left( \phi \right) =f_{0}e^{k\phi }$ and $V\left( \phi \right)
=V_{0}e^{-k\phi }$\cite{D1,DE} we obtain
\begin{equation}
f\left( \phi _{\min }\right) \mathcal{G}=V\left( \phi _{\min }\right) -\frac{%
m_{\phi }^{2}}{k^{2}},  \label{g2}
\end{equation}%
From Eqs. (\ref{acd}, \ref{g1}), we get%
\begin{equation}
V_{eff}(\phi _{\min })=\frac{m_{\phi }^{2}}{k^{2}}.  \label{g3}
\end{equation}%
where $\partial _{\phi }^{2}=\partial ^{2}/\partial \phi ^{2}$\ and $%
V_{eff,\min }$\ as a minimum value of the scalaron effective potential $%
V_{eff}$. The minimum of the potential at $\phi =\phi _{\min }$ should
satisfy $\partial _{\phi }V_{eff}(\phi _{\min })=0$. It is shown that scalar
fields can explain the abundance of dark matter. The scalaron mass change
according to the trace of the energy-momentum tensor \cite{2}. The minimum
of the potential at $\phi =\phi _{\min }$ should satisfy $\partial _{\phi
}V_{eff}(\phi _{\min })=0$, which give a special value of the GB invariant
as
\begin{equation}
\mathcal{G}=-\frac{V_{0}}{f_{0}}e^{-4R_{0}\phi _{\min }}  \label{d3}
\end{equation}%
also we have%
\begin{equation}
V_{eff}(\phi )=V\left( \phi \right) \left[ 1+e^{4R_{0}\left( \phi -\phi
_{\min }\right) }\right]   \label{dd3}
\end{equation}%
If $\phi _{\min }=0$, we get $V_{eff}(\phi )=2V_{0}\cosh \left( 2R_{0}\phi
\right) $. Substituting Eqs.(\ref{va},\ref{d3}) into Eq.(\ref{acd}) then
into Eq.(\ref{d}), the scalaron mass can then be expressed as%
\begin{equation}
m_{\phi }=2\sqrt{2V_{0}}R_{0}e^{-R_{0}\phi _{\min }}.  \label{d4}
\end{equation}%
It is shown that scalar fields can explain the abundance of dark matter. The
scalaron mass change according to the trace of the energy-momentum tensor
\cite{2}. In the large curvature limit \cite{5} we have $R_{0}\phi _{\min }=1
$, one finds
\begin{equation}
R\longrightarrow R_{0}=1/\phi _{\min }=M_{p}^{-2}\rho ,  \label{d5}
\end{equation}%
where $\rho $ is the matter matter-energy. We can then express the scalaron
mass $m_{\phi }$ as a function of the matter-energy density%
\begin{equation}
m_{\phi }=\frac{2\sqrt{2V_{0}}}{eM_{p}^{2}}\rho .  \label{x}
\end{equation}%
The scalaron field $\phi $ becomes dynamical in the low energy density
environment on the cosmological scale. Since the mass $m_{\phi }$ depends on
the matter density $\rho $, thus, the scalaron becomes heavy in the
high-density region of matter. This feature is called the chameleon
mechanism which is one of the screening mechanism in the modified gravity
\cite{15h}. The chameleon mechanisms is defined when the scalaron mass
depends on the environment surrounding the scalaron field. The scalaron $%
\phi $ is regarded as a dynamical dark matter and can be a dark matter
candidate. Similar topics had been researched in many literatures \cite%
{15h,16h}.

\section{Maximally symmetric vacuum solutions}

Recently, there are several papers which study the Particle production
induced by vacuum decay \cite{QQ2}. Following these concepts, here, we will
interpret the Gauss-Bonnet vacua by the production of dark matter particles.
When gravity is taken into consideration, the vacua are those with maximally
symmetric spaces \cite{NN3}. In the maximally symmetric space, the scalar
curvature of de Sitter space is given by\
\begin{equation}
R=\frac{2D}{D-2}\Lambda ,  \label{S1}
\end{equation}%
where $\Lambda $ is the cosmological constant. In the case of the positive $%
\Lambda $, we have the de Sitter solution. In the large limit, we obtain$\
1/\phi _{\min }=\frac{2D}{D-2}\Lambda _{0}$. In $4$-dimensional vacuum, the
equation of motion has a solution that $R_{\mu \nu }=\Lambda g_{\mu \nu }$,
implies that $R=4\Lambda $. The Minkowski spacetime as the vacuum of the $%
\Lambda _{0}=0$. The singularity problem \cite{B4} corresponds to $%
R_{0}\rightarrow \infty $ (the curvature singularity) or $\phi _{\min }$ $=0$%
. From Eq.(\ref{d4}), Eq.(\ref{d5}) and Eq.(\ref{S1}), for $e^{-2R_{0}\phi
_{\min }}\approx 1-2R_{0}\phi _{\min }$ and $R_{0}\longrightarrow R$ one
obtain
\begin{equation}
2\phi _{\min }R^{2}-R+\frac{m_{\phi }}{2\sqrt{2V_{0}}}=0,
\end{equation}%
which can reduce to%
\begin{equation}
4\phi _{\min }\left( \frac{D}{D-2}\right) ^{2}\Lambda ^{2}-\frac{D}{D-2}%
\Lambda +\frac{m_{\phi }}{4\sqrt{2V_{0}}}=0.
\end{equation}%
In maximally symmetric vacuum solutions, there are two branches of solutions
for the effective cosmological constant,%
\begin{equation}
\Lambda _{\pm }=\frac{D-2}{8D\phi _{\min }}\left( 1\pm \sqrt{1-\frac{%
4m_{\phi }}{\sqrt{2V_{0}}}\phi _{\min }}\right) ,  \label{S3}
\end{equation}%
In the limit where $2m_{\phi }\phi _{\min }\ll \sqrt{2V_{0}}$, the two
branches are given by%
\begin{eqnarray}
\Lambda _{+} &=&\frac{D-2}{4D\phi _{\min }}\left( 1-\frac{m_{\phi }\phi
_{\min }}{\sqrt{2V_{0}}}\right) , \\
\Lambda _{-} &=&\frac{D-2}{4D}\frac{m_{\phi }}{\sqrt{2V_{0}}}.
\end{eqnarray}%
Using Eq. (\ref{x}) and $1/\phi _{\min }=\frac{2D}{D-2}\Lambda _{0}$, the
above equations can be further rewritten as%
\begin{equation}
\Lambda _{+}=\frac{\Lambda _{0}}{2}\left( 1-\frac{2\phi _{\min }\rho }{%
eM_{p}^{2}}\right) \text{; \ }\Lambda _{-}=\Lambda _{0}\frac{\phi _{\min
}\rho }{eM_{p}^{2}}.  \label{S4}
\end{equation}%
We notice that $\Lambda _{+}=\frac{\Lambda _{0}}{2}-\Lambda _{-}$. In
particular, the large curvature limit $1/4=\Lambda _{0}\phi _{\min }$, we
have%
\begin{equation}
\frac{\Lambda _{0}}{2}=\Lambda _{+}+\Lambda _{-}\text{; }\Lambda _{-}=\frac{%
\rho }{4eM_{p}^{2}};\text{ \ \ \ }\Lambda _{0}=\frac{1}{4\phi _{\min }}=%
\frac{\rho }{4M_{p}^{2}},
\end{equation}%
or equivalently%
\begin{equation}
\Lambda _{+}=\frac{\left( e-2\right) }{2e}\Lambda _{0}\text{ ;\ }\Lambda
_{-}=\frac{\Lambda _{0}}{e};\text{ \ \ \ }\Lambda _{0}=\frac{\rho }{%
4M_{p}^{2}}  \label{SS4}
\end{equation}%
Let us now comment on the two solutions above. The cosmological constant $%
\Lambda _{+}$ is equivalent to that found by \cite{D2} which is proportional
to the mass of the scalar field (chameleonic dark matter \cite{2}). If $\rho
$ decreases over time, the value of $\Lambda _{+}$ increases to reach $%
\Lambda _{0}$. On the other hand, the second branch $\Lambda _{-}$ depends
on the matter density. We notice that $\Lambda _{+}>0\ $and $\Lambda _{-}>0$%
, instead of studying the false vacuum forming inside the true vacuum in the
bubble geometry \cite{B2}. The two roots $\left( \Lambda _{-},\Lambda
_{+}\right) $ can represent two faces of the same true vacuum. In a vacuum
(without matter), the chameleon mechanism will be zero, which corresponds to
dark energy. The chameleon mechanism appears when there is the matter Eq.(%
\ref{x}), the vacua are those with $\Lambda _{-}$. The vacuum $\Lambda _{-}$%
\ will be spontaneously produced with ordinary matter fields. The
Gauss-Bonnet vacuum becomes a chameleon if there is matter, which solves the
problem of the Gauss-Bonnet vacuum suffering from perturbative ghost
instability \cite{BB}. This justifies that the matter has an impact on the
vacuum, is that the two vacua are separated by a domain wall, composed of
ordinary matter in the thin wall approximation \cite{B2}.\ In particular,
the large curvature limit $1/\phi _{\min }=R_{0}=4\Lambda _{0}$ Eq.(\ref{SS4}%
), must satisfy $\Lambda _{0}=\rho /4M_{p}^{2}$ and $\Lambda _{-}=\rho
/4eM_{p}^{2}$, we can show that $\Lambda _{+}+\Lambda _{-}=\Lambda _{0}$,
which corresponds to $\Lambda _{-}/\Lambda _{0}\approx 0,368\equiv 36,8\%$
and $\Lambda _{+}/\Lambda _{0}\approx 0,132\equiv 13,2\%$. The percentage $%
\Lambda _{-}/\Lambda _{0}\equiv 36,8\%$ is close to the density of matter
and dark matter in the universe \cite{QQ}. It is interesting to note that $%
50\%=\left( 36,8\%+13,2\%\right) $ of space-time gets a mass in a region
occupied by $\left( \Lambda _{-},\Lambda _{+}\right) $. In this case, we
introduce the parameter $X$ which explains the lack of $50\%$ of $\Lambda
_{0}$: $\Lambda _{0}=\Lambda _{+}+\Lambda _{-}+X$, i.e. $X/\Lambda
_{0}=0,5\equiv 50\%$. So, we can associate the parameter $X$ to dark energy.

\begin{table}[H]
\begin{center}
\begin{equation*}
\begin{tabular}{ccc}
\hline
Energy content & parameter & pourcentage \\
Matter+Dark matter & $\Lambda _{-}$ & $\Lambda _{-}/\Lambda _{0}\equiv
36,8\% $ \\
Dark energy & $\Lambda _{+}+X$ & $\left( \Lambda _{+}+X\right) /\Lambda
_{0}=63,2\%$ \\
Universe & $\Lambda _{0}$ & $100\%$ \\ \hline
\end{tabular}%
\end{equation*}%
\end{center}
\caption{Numerical estimate of the percentages of ordinary matter, dark
matter and dark energy in the universe \protect\cite{QQ}.}
\label{t3}
\end{table}
We notice that the dark energy is described by the parameter $\Lambda _{+}$
and an unknown parameter $X$ table (\ref{t1}).

\section{Functional coupling and Barrow entropy}

In this section, we explain in detail how to construct the equation of
motion of the Einstein-Gauss-Bonnet gravity. We begin by reviewing the the
4D EsGB action%
\begin{equation}
\mathcal{S}=\frac{M_{p}^{2}}{2}\int d^{4}x\sqrt{-g}\left( R+f\left( \phi
\right) \mathcal{G}-\frac{1}{2}g^{\mu \nu }\partial _{\mu }\phi \partial
_{\nu }\phi -V\left( \phi \right) \right) +\mathcal{S}_{m},  \notag
\label{0F1}
\end{equation}%
where $M_{p}=1/\left( 8\pi G_{N}\right) =1.221\times 10^{19}GeV$ is the
reduced Planck mass, $R$ is the Ricci scalar, $S_{m}$ is the matter action
and $f\left( \phi \right) $ is a functional coupling of the scalar field $%
\phi $. In the above equation $\left( \mu ,\nu \right) =\left(
0,1,2,3\right) $. We define the GB term as
\begin{equation}
\mathcal{G}\equiv R^{2}-4R_{\mu \nu }R^{\mu \nu }+R_{\mu \nu \rho \sigma
}R^{\mu \nu \rho \sigma }.  \label{F2}
\end{equation}%
The Kretschmann scalar is $R_{\mu \nu \rho \sigma }R^{\mu \nu \rho \sigma }$%
. The variation with respect to the field $\phi $ gives us the equation of
motion for the scalar field%
\begin{equation}
\square \phi =\partial _{\phi }V_{eff}\left( \phi \right) ,
\end{equation}%
where $\square \equiv \nabla _{\mu }\nabla ^{\mu }$ and the effective
potential is%
\begin{equation}
V_{eff}\left( \phi \right) =V\left( \phi \right) -f\left( \phi \right)
\mathcal{G}.  \label{F4}
\end{equation}%
Varying the action (\ref{F1}) over the metric $g_{\mu \nu }$, we obtain the
following equations of motion:%
\begin{equation}
G^{\mu \nu }+\mathcal{K}^{\mu \nu }+f\left( \phi \right) \mathcal{H}^{\mu
\nu }+\frac{1}{2}\left[ \mathcal{T}_{\phi }^{\mu \nu }-g^{\mu \nu
}V_{eff}\left( \phi \right) \right] =\frac{1}{2}\kappa ^{2}T^{\mu \nu },
\notag  \label{0EM}
\end{equation}%
where the Einstein tensor is $G^{\mu \nu }=R^{\mu \nu }-\frac{1}{2}g^{\mu
\nu }R$, the matter stress tensor is $T^{\mu \nu }=-\frac{2}{\sqrt{-g}}\frac{%
\delta \mathcal{S}_{m}}{\delta g_{\mu \nu }}$. On the other hand the $%
\mathcal{K}^{\mu \nu }$ and $\mathcal{H}^{\mu \nu }$ are given by%
\begin{eqnarray}
\mathcal{K}^{\mu \nu } &=&4[G^{\mu \nu }\square +\frac{1}{2}R\nabla ^{\mu
}\nabla ^{\nu }+\left( g^{\mu \nu }R^{\rho \sigma }-R^{\mu \rho \nu \sigma
}\right) \nabla _{\rho }\nabla _{\sigma },  \notag  \label{0F6} \\
&&-R^{\nu \rho }\nabla _{\rho }\nabla ^{\mu }+R^{\mu \rho }\nabla _{\rho
}\nabla ^{\nu }]f\left( \phi \right)  \label{F6}
\end{eqnarray}%
\begin{eqnarray}
\mathcal{H}^{\mu \nu } &=&2R^{\mu \rho \sigma \tau }R_{\text{ \ }\rho \sigma
\tau }^{\nu }-RR^{\mu \nu }  \notag  \label{0F7} \\
&&+\frac{1}{2}R_{\text{ \ }\rho }^{\mu }R^{\nu \rho }-R^{\mu \rho \sigma
\tau }R_{\text{ \ }\rho \sigma \tau }^{\nu }.  \label{F7}
\end{eqnarray}%
The tensor $\mathcal{K}^{\mu \nu }$ represents an operator which acts on $%
f\left( \phi \right) $. The energy-momentum tensor for the scalar field is%
\begin{equation}
\mathcal{T}_{\phi }^{\mu \nu }=\nabla ^{\mu }\phi \nabla ^{\nu }\phi -\frac{1%
}{2}g^{\mu \nu }\nabla _{\rho }\phi \nabla ^{\rho }\phi .  \label{F8}
\end{equation}%
The stress tensor for anisotropic compact object is given as%
\begin{equation}
T^{\mu \nu }=\left( \rho +P_{t}\right) u^{\mu }u^{\nu }+P_{t}g^{\mu \nu
}+\left( P-P_{t}\right) \chi ^{\mu }\chi ^{\nu },  \label{F9}
\end{equation}%
with energy density $\rho _{CO}=\rho (r)\equiv c^{2}\rho (r)$, transverse
pressure $P_{t}(r)$ and radial pressure $P(r)$ of\ the homogeneously
distributed matter in the compact object (CO), where $u^{\mu }$ is the
four-velocity of the fluid and $\chi ^{\mu }$ is the unit space-like vector
in the radial direction.\newline
In limit $\alpha \longrightarrow 0$, one can see that this is equivalent to
the original form given by the Bekenstein-Hawking entropy. One can check
that for $D=2$, the topological term characterized by a vanishes Lovelock
coupling $\alpha $. The Barrow entropy \cite{PLB1} is a new black hole
entropy which is given by
\begin{equation}
S=\pi \left( \frac{A}{A_{0}}\right) ^{1+\delta /2},
\end{equation}%
where $0\leq \delta \leq 1$, $A$ is the black hole horizon area and $A_{0}$
is the Planck area \cite{MBOU3}. When $\delta =0$, the area law is restored,
i.e. $S=\frac{A}{4G}$ $\left( \text{where }A_{0}=4G\right) $ while $\delta
=1 $ represents the most intricate and fractal structure of the horizon.%
\newline
The form the functional $f\left( \phi \right) $ can take $f\left( \phi
\right) =e^{-\gamma \phi }$ \cite{L1}, where $\gamma $ is a constant, which
corresponds to EGB gravity coupled with dilaton that arises as a low-energy
limit of the string theory \cite{L2}. Using the same principle as this
entropy, we assume that for the compact object (or black hole)\ exterior,
there is a presence of scalar fields $\phi $, while inside the compact
object is replaced by the Gauss-Bonnet (GB) coupling $\alpha $ as
\begin{equation}
\begin{array}{c}
\text{coupling constant=}f\left( \phi \right) ,\text{ \ \ CO\ exterior }%
\left( D=4\right) \\
\text{coupling constant=}\frac{\alpha }{D-4},\text{ \ \ CO\ interior }\left(
D\rightarrow 4\right)%
\end{array}%
.  \label{F11}
\end{equation}%
The GB coupling $\alpha $ is measured in\textbf{\ }$km^{2}$. In the CO\
exterior\ interior, we have rescaled the coupling constant $\alpha
\rightarrow \alpha /\left( D-4\right) $. The negative (positive) $\alpha $
leads to a decrease (increase) of the CO radius and the maximum mass \cite%
{L5}. If $\alpha <0$\ the solution is still the anti-de Sitter (AdS) space,
if $\alpha >0$\ the solution is the de Sitter (dS) space \cite{L3}. We
investigate in detail the impact of the Gauss-Bonnet coupling on the
properties of an anisotropic compact object, such as mass, radius and the
factor of compactness. Considering the limit $D\rightarrow 4$, and it has an
effect on gravitational dynamics in 4D. Additionally, at the CO boundary $%
(r=R)$, the GB coupling must be continuous, i.e. $f\left( \phi \right)
\equiv \frac{\alpha }{D-4}$. On the other hand, the function $f\left( \phi
\right) $ describes the star's exterior region. To study the equations of
motion inside and outside CO, we differentiate between two cases:\newline
In CO\ interior ($D\rightarrow 4$) we have:
\begin{equation}
G^{\mu \nu }+\alpha \left( \mathcal{H}^{\mu \nu }+\frac{1}{2}g^{\mu \nu }%
\mathcal{G}\right) =\frac{\kappa ^{2}}{2}T^{\mu \nu }.  \label{F12}
\end{equation}%
In CO\ exterior ($D=4$) we have:%
\begin{equation}
G^{\mu \nu }+\mathcal{K}^{\mu \nu }+f\left( \phi \right) \left[ \mathcal{H}%
^{\mu \nu }-\frac{1}{2f\left( \phi \right) }g^{\mu \nu }\left( \nabla
_{\lambda }\phi \nabla ^{\lambda }\phi +4V_{eff}\left( \phi \right) \right) %
\right] =\frac{1}{2}\kappa ^{2}T^{\mu \nu },  \label{F13}
\end{equation}%
with $g_{\mu \nu }\mathcal{T}_{\phi }^{\mu \nu }=-\nabla _{\lambda }\phi
\nabla ^{\lambda }\phi $. In CO\ interior ($D\rightarrow 4$) we have:\ The
GB invariant can be greatly simplified to the matter density \cite{DM} and
using Eq.(\ref{F4}) we obtain%
\begin{equation}
\begin{array}{c}
\rho _{f(\phi )}=4\mathcal{G}-\frac{4}{f\left( \phi \right) }\left( \frac{1}{%
4}\nabla _{\lambda }\phi \nabla ^{\lambda }\phi +V\left( \phi \right)
\right) ,\text{ \ \ CO\ exterior }\left( D=4\right) \\
\rho _{CO}=\mathcal{G},\text{ \ \ \ \ \ \ \ \ \ \ \ \ \ \ \ \ \ \ \ \ \ \ \
\ \ \ \ \ \ \ \ \ \ \ \ \ \ \ \ \ \ \ \ \ \ \ \ \ \ CO\ interior }\left(
D\rightarrow 4\right)%
\end{array}%
.  \label{F14}
\end{equation}%
In this case, the term $\rho _{CO}$ represents the density of matter in
compact objects, and $\rho _{f(\phi )}$ is the density of dark matter
surrounding these objects. Note that the relation between $\rho _{f(\phi )}$
and $\rho _{CO}$ highlight the chameleon dark matter \cite{L12,L13}.
Formally, at the points where the dark matter density $\rho _{DM}$ equal to $%
\rho _{f(\phi )}$. For $V\left( \phi \right) \approx -\frac{1}{4}\nabla
_{\lambda }\phi \nabla ^{\lambda }\phi $, we obtain $\rho _{DM}\approx 4\rho
_{CO}$, which is in good agreement with the observation data of the
percentages of dark matter and the matter in the Universe \cite{L16,MBOU4}: $%
\rho _{DM}\equiv 80\%$ and $\rho _{matter}(\rho _{CO})\equiv 20\%$. \newline
Next, we assume that $\phi =\phi (t)$. In cosmological and quintessence
behavior \cite{L14,L15}, the energy density $\rho _{\phi }$ and pressure $%
P_{\phi }$ of the scalar field are given by%
\begin{eqnarray}
\rho _{\phi } &=&\frac{1}{2}\dot{\phi}^{2}+V\left( \phi \right) ,
\label{F15} \\
P_{\phi } &=&\frac{1}{2}\dot{\phi}^{2}-V\left( \phi \right) .  \notag
\end{eqnarray}%
The quintessence models describe dark energy with a scalar field $\phi $. In
this case, $\rho _{\phi }$ and $P_{\phi }$ are respectively, the density and
the pressure of the dark energy (DE). The Planck Collaboration \cite{L16}
provides constraints on the equation of state $\omega _{\phi }=P_{\phi
}/\rho _{\phi }\approx -1.028\pm 0.032$. Starting from Eqs. (\ref{F15}), we
obtain%
\begin{equation}
\rho _{DM}=4\rho _{CO}+\frac{1}{f\left( \phi \right) }\left( P_{\phi }-3\rho
_{\phi }\right) .  \label{F16}
\end{equation}%
For small $f\left( \phi \right) $, the DM chameleon effect vanishes and the
DM density depends only on $\left( P_{\phi }-3\rho _{\phi }\right) /f\left(
\phi \right) $. \newline
The scalar field always sits at the minimum of its effective potential. We
assume that a massive scalar field begins oscillating about a minimum. The
mass of small fluctuations around $\phi _{\min }$ gives a new scalar field
mass as effective mass by $m_{eff}^{2}=\left. \frac{\partial ^{2}}{\partial
\phi ^{2}}V_{eff}(\phi )\right\vert _{\phi =\phi _{\min }}$. From Eqs. (\ref%
{F4},\ref{F15},\ref{F16}) we obtain%
\begin{equation}
m_{eff}^{2}=\frac{1}{2}\frac{\partial ^{2}}{\partial \phi ^{2}}\left( \rho
_{\phi }-P_{\phi }+\frac{\rho _{CO}}{\rho _{DM}-4\rho _{CO}}\left( 3\rho
_{\phi }-P_{\phi }\right) \right) _{\phi =\phi _{\min }}.  \label{F17}
\end{equation}%
From Eq. (\ref{F16}), the functional coupling is given by%
\begin{equation}
f\left( \phi \right) =\frac{P_{\phi }-3\rho _{\phi }}{\rho _{DM}-4\rho _{CO}}%
.  \label{F}
\end{equation}%
In the void ($\rho _{CO}=0,\omega _{\phi }\sim -1$), we have $f\left( \phi
\right) =-4\rho _{\phi }/\rho _{DM}$. In the CO surface, we assume that $%
\rho _{DM}\approx 0$ and $P_{\phi }\approx 0$, so we get $f_{surface}\left(
\phi \right) \approx 3\rho _{\phi }/4\rho _{CO}$. Since $\rho _{\phi }$
represents the density of DE according to quintessence, the effect of DE is
weak on the CO surface, which shows that $f_{surface}\left( \phi \right)
\left( \propto \rho _{\phi }\right) \rightarrow 0$. Inside CO, we have ($%
P_{\phi }=\rho _{\phi }=0$), i.e. $f\left( \phi \right) =0$, which exactly
corresponds with the assumption Eq. (\ref{F11}). For this reason, we exclude
$f\left( \phi \right) $ inside matter, and we replace it with the GB
coupling $\alpha $ (see the next section).

\section{EGB primordial black holes}

We start by the action of Einstein-Gauss-Bonnet (EGB) gravity \cite{D3} in $%
4 $-dimensions as

\begin{equation}
\mathcal{I}=\int d^{D}x\sqrt{-g}\left( \frac{M_{p}^{2}}{2}R-2\Lambda
+f\left( \phi \right) \left( R^{2}-4R_{\mu \nu }R^{\mu \nu }+R_{\mu \nu \rho
\sigma }R^{\mu \nu \rho \sigma }\right) \right) ,  \label{a1}
\end{equation}%
where $S_{m}$ is the matter action, $M_{p}^{2}=\frac{c^{4}}{8\pi G}\approx
2\times 10^{18}\left[ GeV\right] $, $R$ is the Ricci scalar. We define the
Gauss-Bonnet invariant as $\mathcal{G}=\mathcal{L}_{2}=R^{2}-4R_{\mu \nu
}R^{\mu \nu }+R_{\mu \nu \rho \sigma }R^{\mu \nu \rho \sigma },$ where $%
\alpha _{2}=\alpha /\left( D-4\right) $ is the Gauss-Bonnet coupling have
dimensions of $\left[ length\right] ^{2}$, that represent ultraviolet (UV)
corrections to Einstein theory. To solving the field equation we obtain the
black hole solution $ds^{2}=-f(r)dt^{2}+\frac{1}{f(r)}dr^{2}+r^{2}\left(
d\theta ^{2}+\sin ^{2}\theta d\phi ^{2}\right) $. Taking the limit $%
D\longrightarrow 4$, we obtain the exact solution in closed form
\begin{equation}
f(r)\approx 1+\frac{r^{2}}{2\alpha }\left( 1-\sqrt{1+4\alpha \left( \frac{2M%
}{r^{3}}\pm \frac{1}{l^{2}}\right) }\right) .  \label{a3}
\end{equation}%
This last solution could be obtained directly from the derivation done in
\cite{D8}. In the limit $r\longrightarrow \infty $ with vanishing black hole
charge, we asymptotically obtain the GR Schwarzschild solution. In the limit
$\alpha \longrightarrow 0$, we can recover the Reissner-Nordstr\"{o}m-AdS
solution. If $\alpha <0$\ the solution is still an AdS space, if $\alpha >0$%
\ the solution is a de Sitter (dS) space, \cite{D7}. The solutions show that
the event horizon is located at $R_{\pm }=M\pm \sqrt{M^{2}-\alpha },$ where $%
R_{H}=R_{+}$ and $R_{-}$ are the event horizon and the Cauchy horizon radius
of the EGB black hole \cite{D0}. We can express the ADM mass $M$ of the
black hole in terms of $R_{H}$ by solving $f(r)=0$ for $r=R_{H}$ resulting in%
\begin{equation}
M=\frac{l^{-2}R_{H}^{4}+R_{H}^{2}+\alpha }{2R_{H}}.  \label{a5}
\end{equation}%
The Hawking temperature of the EBG black hole can be calculated as

\begin{equation}
T=\frac{3l^{-2}R_{H}^{4}+R_{H}^{2}-\alpha }{8\pi \alpha R_{H}+4\pi R_{H}^{3}}%
.  \label{a7}
\end{equation}%
The thermodynamic volume $V=\frac{4\pi R_{H}^{3}}{3}$ is the conjugate
variable to the pressure. The parameters $V$ and $A$ are the conjugate
quantities of the pressure $P$ and GB coupling parameter $\alpha $. The
event horizon in spacetime can be located by solving the metric equation: $%
f(r)=0$, and from Eq .(\ref{a3}) we obtain
\begin{equation}
\rho _{BH}+P=\frac{3}{8\pi R_{H}^{2}}+\frac{3\alpha }{8\pi R_{H}^{4}},
\label{b6}
\end{equation}%
where $M=\rho _{BH}V$, $V=4\pi r^{3}/3$ and $P=3/8\pi l^{2}$. For the limit $%
\alpha \longrightarrow 0$, we can recover the density $\rho _{\Lambda
}=3/16\pi R_{H}^{2}$ of the holographic dark energy (DE) \cite{D9}. In the
spatially flat homogeneous and isotropic universe in the FLRW, the
(modified) Friedmann equations can be obtained \cite{D3} $H^{2}+\alpha H^{4}=%
\frac{8\pi G}{3}\rho +\frac{\Lambda }{3}$ and $\left( H^{2}+\alpha
H^{4}\right) \dot{H}=-4\pi G\left( \rho +P\right) $. This equation looks
like the Eq .(\ref{b6}), i.e. can describe the dynamics in space-time
associated to a black hole. For Gauss--Bonnet branch, we introduce the
topological density as%
\begin{equation}
\rho _{\alpha }=\rho _{\alpha }^{\left( 2\right) }=\frac{\alpha }{16\pi
R_{H}^{4}}.  \label{b7}
\end{equation}%
From Eq. (\ref{b6}) we obtain the Van der Waals equation:%
\begin{equation}
\left( P+\rho _{BH}\right) \left( V-\frac{4\pi \alpha R_{H}}{3}\right) =%
\frac{R_{H}}{2}.  \label{b8}
\end{equation}%
In the limit $\alpha \longrightarrow 0$ and $P\gg \rho $, we can recover the
ideal gas law. The critical point occurs when $P(V)$. We note that the
Gauss-Bonnet coupling $\alpha $\ represente measure of the average
attraction between particles. Before proceeding further, we note that the
mass $M$ can be interpreted as a chemical enthalpy \cite{R9}, which is the
total energy of the black hole \cite{R10} including both the energy $PV$ and
the internal energy $E$ and required to displace the vacuum energy of its
environment. The Hubble horizon mass is connected with the Smarr formula as%
\begin{equation}
M_{H}+PV=\frac{R_{H}}{2}+\frac{\alpha }{2R_{H}},  \label{bb1}
\end{equation}%
where $M_{H}=\frac{4\pi R_{H}^{3}}{3}\rho $\ is the Hubble horizon mass. We
have also
\begin{equation}
R_{H}\approx 2M-\frac{\alpha }{2M}.  \label{bb2}
\end{equation}%
We can recover the Schwarzschild radius for $\alpha \approx 0$. This
expression gives an interpretation of the difference between Schwarzschild
and EGB black holes. From Eqs. (\ref{bb1},\ref{bb2}), we obtain the the
Smarr formula:%
\begin{equation}
M=M_{H}+PV+\frac{\alpha }{4M}\left( 1-\frac{1}{1-\frac{\alpha }{4M^{2}}}%
\right) .  \label{bb3}
\end{equation}%
In the limit $\alpha \longrightarrow 0,$ we can recover the standard Smarr
formula. Let us mention that for the AdS-Schwarzschild limit and using $%
M\approx M_{H}$, we get $PV\propto =-\alpha /4M\propto T$. This expression
gives an interpretation to the term that depends on $\alpha $, as a solution
of AdS-GB gravity in the presence of a perfect fluid \cite{D4}. This is not
valid for the dS-GB black holes.
\begin{equation}
PV=\left( M-M_{H}\right) -\frac{\alpha }{4M}\left( 1-\frac{1}{1-\frac{\alpha
}{4M^{2}}}\right) .  \label{bb4}
\end{equation}%
For Gauss-Bonnet branch we obtain $\left( 1-\frac{4M^{2}}{\alpha }\right)
\left( 1-\frac{\alpha }{4M^{2}}\right) =1$. One of two solutions to this
equation is the Golden ratio ($\alpha >0$) in de Sitter (dS) space: $\frac{%
\alpha }{4M^{2}}=\frac{1+\sqrt{5}}{2}$ or\ $\frac{4M^{2}}{\alpha }=\frac{1+%
\sqrt{5}}{2}$. If $\alpha <0$\ the solution is still an AdS space: $\frac{%
\alpha }{4M^{2}}=\frac{1-\sqrt{5}}{2}$ or $\frac{4M^{2}}{\alpha }=\frac{1-%
\sqrt{5}}{2}.$ We obtain the spectrum of mass in the Gauss-Bonnet branch is%
\begin{equation}
M_{\alpha }=\frac{\sqrt{\gamma \alpha }}{2},  \label{bb7}
\end{equation}%
with $\gamma =\left\{ \frac{2}{1-\sqrt{5}},\frac{1-\sqrt{5}}{2},\frac{2}{1+%
\sqrt{5}},\frac{1+\sqrt{5}}{2}\right\} $. It can be written as $\gamma
\approx \left\{ -1.618,-0.618,0.618,1.618\right\} $. This shows that the
Gauss-Bonnet coupling represents a topological black hole mass.
\begin{table}[H]
\begin{equation*}
\begin{tabular}{ccc}
\hline
$\alpha $ & $\gamma $ & $M_{\alpha }$ \\ \hline\hline
$-0.5$ & $-1.618$ & $0.449$ \\
$-0.5$ & $-0.618$ & $0.277$ \\
$0.5$ & $0.618$ & $0.277$ \\
$0.5$ & $1.618$ & $0.449$ \\ \hline
\end{tabular}%
\end{equation*}%
\caption{Numerical estimate of the values of the PBH mass, according to the
frequency $f$ and the interval of $N$ between $10^{-47}$ and $60$. Such as $%
f=2.561\times 10^{33}\frac{N}{M}.$}
\label{t1}
\end{table}
Taking the derivative with respect to time $t$ in both sides of Eq. (\ref%
{bb2}), thus the mass loss rate of a black hole is obtained as%
\begin{equation}
\frac{dM}{dt}\approx \frac{\dot{R}_{H}}{2}\left( 1-\frac{\alpha }{2M^{2}}%
\right) ,  \label{bb8}
\end{equation}%
with $\dot{R}_{H}=\frac{dR_{H}}{dt}$. Using the 4D Stefan-Boltzmann law \cite%
{Z18,Z19} $\frac{dM}{dt}=-\pi ^{2}AT^{4}/60$, with $T$ and $A$\ are
temperature and area of black hole, respectively. Evidently, we have $\alpha
/M^{2}\propto T^{4}$. For the Gauss-Bonnet branch and from Eq. (\ref{bb7}),
we have $\frac{dM}{dt}\approx \frac{\dot{R}_{H}}{2}\left( 1-2\gamma \right) $%
. Taking the Friedmann equation $HR_{H}=1$ \cite{D9} and from Eq .(\ref{bb2}%
) we have
\begin{equation}
M_{\pm }(t)=\frac{1}{4H\left( t\right) }\pm \frac{1}{4}\left( \frac{1}{%
H\left( t\right) }+4\alpha \right) ^{1/2}.  \label{bb9}
\end{equation}%
Focusing on $M_{-}(t)$, one can find the condition $\frac{1}{H}\left( \frac{1%
}{H}-1\right) \geq \frac{4\alpha }{\theta ^{2}}$.
\begin{figure}[]
\centering\includegraphics[width=15cm]{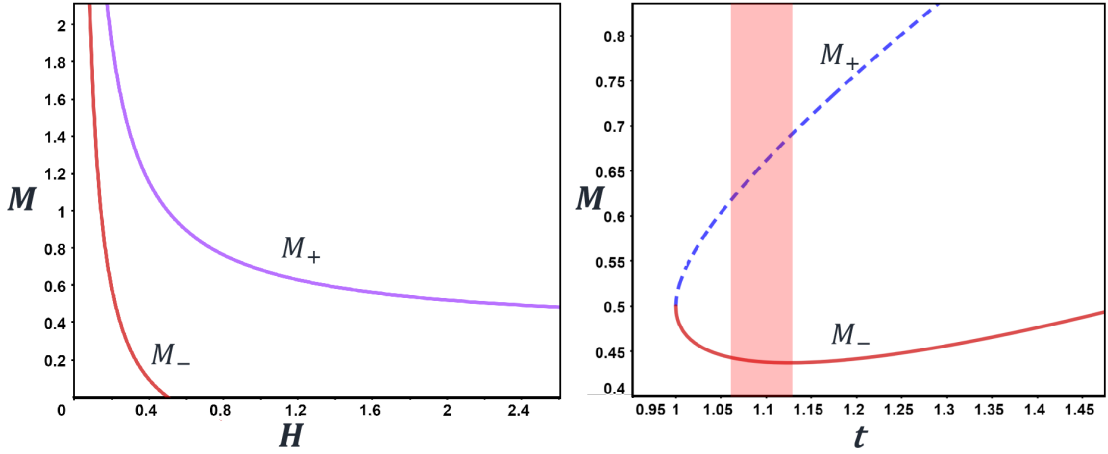}
\caption{Left, curves of $M_{\pm }$ vs $H$ from Eq.\ (\protect\ref{bb9})with
$\protect\alpha =0.5$. Right, $M_{+}$ vs $t$\ (dashed line) and $M_{-}$ vs $%
t $\ (solid line) for the values $q=1/2$ and $\protect\alpha =-0.5$ .}
\label{F3}
\end{figure}
For $\alpha =0.5$ we plot $M_{\pm }$ vs $H$ Fig. (\ref{F3}, left). Using $%
H=q/t$ in the radiation-dominated era ($q=1/2$) for $\alpha =-0.5$, we plot $%
M_{\pm }$ vs $t$ Fig. (\ref{F3}, right). In Fig. (\ref{F3}, right), the
domain in red, represents the interval where the evolution of the mass with
respect to time begins to increase. While, before this interval, the mass
decreases over time which is due to Hawking evaporation. According to this
figure, two types of PBHs have opposite properties such as the evolution of
their masses over time.

\section{Conclusion}

The coupling between the scalar field $\phi $ and the 4D
Einstein-Gauss-Bonnet$\ $gravity action is studied. We have studied the
model describing the compact stars in 4D Einstein-Gauss-Bonnet gravity
surrounded by scalar dark matter. We have made a comparison between the star
interior and the exterior region. In this case, the Gauss-Bonnet coupling
describes the interior structure of the black hole, while the coupling
function $f\left( \phi \right) $ describes the black hole exterior region.
We have shown that $f\left( \phi \right) =0$ inside the black hole. By
solving the equations of motion based on the chameleon mechanism. We refer
to $\phi $ as a chameleon field, since its physical properties, such as its
mass, depend sensitively on the environment. It can be used to demonstrate
the relation between the cosmological constant and the matter density. The
Gauss-Bonnet vacuum have a chameleon structure, and the mass $m_{\phi }$
appears with a small fluctuations. Moreover, we have studied the vacua in
maximally symmetric solutions, and we obtain two branches of solutions, i.e.
there are two such vacua. The first solution corresponds to the dark energy
and the second represents the chameleonic vacua or dark matter. We have
found strong mixing between the two vacua, which is presented by the
chameleon mechanism. We also discussed the equation of state parameter in
the model of the scalar field $\phi $ minimally coupled to EGB gravity, also
in the maximally symmetric space. The percentages of the effective
cosmological constants are in good agreement with those of dark matter and
dark energy in the Universe.\newline
We have considered 4D Einstein-Gauss-Bonnet gravity. We establish the
relationship between a topological mass spectrum and the golden ratio. We
have obtained the mass loss rate, which gives by taking the Friedmann
equation of type $HR_{H}=1$ \cite{D9}, the evolution of the mass over time,
and the evaporation Hawking time. For such a choice and fixed values of
Lovelock coupling, we have plotted the mass of PBH versus the Hubble
parameter and time.


\begin{thebibliography}{99}
\bibitem{0000} \textrm{\ }Del Popolo, Antonino, and Morgan Le Delliou.
"Small scale problems of the $\Lambda $CDM model: a short review." Galaxies
5.1 (2017): 17.

\bibitem{G1} D. Glavan, \& C. Lin, Einstein-Gauss-Bonnet Gravity in
Four-Dimensional Spacetime, Phys. Rev. Lett. 124(8) (2020) , 081301. \textrm{%
\ }

\bibitem{DE} Easson, D. A., Manton, T., \& Svesko, A. (2020). D$%
\longrightarrow $4 Einstein-Gauss-Bonnet gravity and beyond. Journal of
Cosmology and Astroparticle Physics, 2020(10), 026.

\bibitem{R1} D. Lovelock, The four-dimensionality of space and the Einstein
tensor, J. Math. Phys. 13(6) (1972), 874-876.

\bibitem{F1} Bonifacio, J., Hinterbichler, K., \& Johnson, L. A. (2020),
Amplitudes and 4D Gauss-Bonnet Theory, Phys. Rev. D, 102(2), 024029.

\bibitem{F2} G\"{u}rses, M., \c{S}i\c{s}man, T. \c{C}., \& Tekin, B. (2020).
Comment on \textquotedblleft Einstein-Gauss-Bonnet Gravity in
Four-Dimensional Spacetime\textquotedblright , Phys. Rev. Lett., 125(14),
149001.

\bibitem{F3} Wang, D., \& Mota, D. (2021). 4D Gauss--Bonnet gravity:
Cosmological constraints, H0 tension and large scale structure, Phys. Dark
Universe, 100813.

\bibitem{F4} Wu, C. H., Hu, Y. P., \& Xu, H. (2021). Hawking evaporation of
Einstein--Gauss--Bonnet AdS black holes in D$\geq $\ 4 dimensions, Eur.
Phys. J. C, 81(4), 1-9.

\bibitem{N0} Horndeski, G. W. (1974). Second-order scalar-tensor field
equations in a four-dimensional space. Int. J. Theor. Phys., 10(6), 363-384.

\bibitem{N1} Guo, M., \& Li, P. C. (2020), Innermost stable circular orbit
and shadow of the 4 D Einstein--Gauss--Bonnet black hole, Eur. Phys. J. C,
80(6), 1-8.

\bibitem{N9} Konoplya, R. A., \& Zinhailo, A. F., 2020, \textrm{Eur. Phys.
J. C,}. 80(11), 1-13.

\bibitem{N10} Banerjee, A., Tangphati, T., \& Channuie, P., 2021, \textrm{%
Astrophys. J.}, 909(1), 14.

\bibitem{N11} Hansraj, S., Banerjee, A., Moodly, L., \& Jasim, M. K., 2020,
\textrm{Class. Quantum Gravity}, 38(3), 035002.

\bibitem{N12} Banerjee, A., Tangphati, T., Samart, D., \& Channuie, P.,
2021, \textrm{Astrophys. J.}, 906(2), 114.

\bibitem{N13} Banerjee, A., \& Singh, K. N., 2021, \textrm{Phys. Dark
Universe}, 31, 100792.

\bibitem{N14} Tangphati, T., Pradhan, A., Errehymy, A., \& Banerjee, A.,
2021, \textrm{Phys. Lett. B}, 136423.

\bibitem{NN1} East, W. E., \& Ripley, J. L. (2021). Evolution of
Einstein-scalar-Gauss-Bonnet gravity using a modified harmonic formulation.
Physical Review D, 103(4), 044040.

\bibitem{NN2} Horndeski, G. W. (1974). Second-order scalar-tensor field
equations in a four-dimensional space. International Journal of Theoretical
Physics, 10(6), 363-384.

\bibitem{BB} Charmousis, C., \& Padilla, A. (2008). The instability of vacua
in Gauss-Bonnet gravity. Journal of High Energy Physics, 2008(12), 038.

\bibitem{B3} Ma, L., \& L\"{u}, H. (2020). Vacua and exact solutions in
lower-D limits of EGB. The European Physical Journal C, 80(12), 1-10.

\bibitem{133} Panah, B. E., Jafarzade, K., \& Hendi, S. H. (2020). Charged
4D Einstein-Gauss-Bonnet-AdS black holes: shadow, energy emission,
deflection angle and heat engine. Nuclear Physics B, 961, 115269.

\bibitem{NN3} Markkanen, T., Nurmi, S., Rajantie, A., \& Stopyra, S. (2018).
The 1-loop effective potential for the Standard Model in curved spacetime.
Journal of High Energy Physics, 2018(6), 1-40.

\bibitem{NN4} Linde, A. D. (1983). Decay of the false vacuum at finite
temperature. Nuclear Physics B, 216(2), 421-445.

\bibitem{NN5} Coleman, S., \& Luccia, F. D. (1980). Gravitational effects on
and of vacuum decay. In EUCLIDEAN QUANTUM GRAVITY (pp. 295-305).

\bibitem{D1} Nojiri, S. I., Odintsov, S. D., \& Sasaki, M. (2005).
Gauss-Bonnet dark energy. Physical Review D, 71(12), 123509.

\bibitem{D2} Glavan, D., \& Lin, C. (2020). Einstein-Gauss-Bonnet Gravity in
Four-Dimensional Spacetime. Physical Review Letters, 124(8), 081301.

\bibitem{MBOU} Bousder, M., Salmani, E., El Fatimy, A., \& Ez-Zahraouy, H.
(2023). Holographic dark energy satisfying the energy conditions in Lovelock
gravity. Annals of Physics, 169282.

\bibitem{QW1} Cat\`{a}, O., Ibarra, A., \& Ingenh\"{u}tt, S. (2016). Dark
matter decays from nonminimal coupling to gravity. Physical review letters,
117(2), 021302.

\bibitem{G2} Aoki, K., Gorji, M. A., \& Mukohyama, S. (2020). A consistent
theory of $D\longrightarrow 4$ Einstein-Gauss-Bonnet gravity. Physics
Letters B, 135843.

\bibitem{rip} Caldwell, R. R., Kamionkowski, M., \& Weinberg, N. N. (2003).
Phantom energy: dark energy with w\TEXTsymbol{<}- 1 causes a cosmic
doomsday. Physical Review Letters, 91(7), 071301.

\bibitem{40} Pi, Shi, et al. "Scalaron from R2-gravity as a heavy field."
Journal of Cosmology and Astroparticle Physics 2018.05 (2018): 042.

\bibitem{30} Khoury, Justin, and Amanda Weltman. "Chameleon cosmology."
Physical Review D 69.4 (2004): 044026.

\bibitem{2} Katsuragawa, Taishi, and Shinya Matsuzaki. "Cosmic history of
chameleonic dark matter in F (R) gravity." Physical Review D 97.6 (2018):
064037.

\bibitem{5} Gannouji, Radouane, M. Sami, and I. Thongkool. "Generic f (R)
theories and classicality of their scalarons." Physics Letters B 716.2
(2012): 255-259.

\bibitem{15h} Katsuragawa, T., \& Matsuzaki, S. (2018). Cosmic history of
chameleonic dark matter in F (R) gravity. Physical Review D, 97(6), 064037.

\bibitem{16h} Choudhury, S., Sen, M., \& Sadhukhan, S. (2016). Can dark
matter be an artifact of extended theories of gravity?. The European
Physical Journal C, 76(9), 1-24.

\bibitem{B4} Frolov, A. V. (2008). Singularity problem with f (R) models for
dark energy. Physical review letters, 101(6), 061103.

\bibitem{B5} Crisostomo, J., Troncoso, R., \& Zanelli, J. (2000). Black hole
scan. Physical Review D, 62(8), 084013.

\bibitem{B2} Shu, F. W. (2020). Vacua in novel 4D Einstein-Gauss-Bonnet
gravity: Pathology and instability?. Physics Letters B, 811, 135907.

\bibitem{QW2} Ema, Y., Jinno, R., \& Nakayama, K. (2020). High-frequency
graviton from inflaton oscillation. Journal of Cosmology and Astroparticle
Physics, 2020(09), 015.

\bibitem{C1} Elder, B., Khoury, J., Haslinger, P., Jaffe, M., M\"{u}ller,
H., \& Hamilton, P. (2016). Chameleon dark energy and atom interferometry.
Physical Review D, 94(4), 044051.

\bibitem{C2} Brax, P., van de Bruck, C., Davis, A. C., Khoury, J., \&
Weltman, A. (2004). Detecting dark energy in orbit: The cosmological
chameleon. Physical Review D, 70(12), 123518.

\bibitem{QQ} Aghanim, N., Akrami, Y., Ashdown, M., Aumont, J., Baccigalupi,
C., Ballardini, M., ... \& Roudier, G. (2020). Planck 2018 results-VI.
Cosmological parameters. Astronomy \& Astrophysics, 641, A6.

\bibitem{QQ1} Garc\'{\i}a-Aspeitia, M. A., \& Hern\'{a}ndez-Almada, A.
(2021). Einstein--Gauss--Bonnet gravity: Is it compatible with modern
cosmology?. Physics of the Dark Universe, 100799.

\bibitem{QQ2} Hashiba, S., Yamada, Y., \& Yokoyama, J. I. (2021). Particle
production induced by vacuum decay in real time dynamics. Physical Review D,
103(4), 045006.

\bibitem{QQ3} Krishnan, C., Mohayaee, R., Colg\'{a}in, E. \'{O}.,
Sheikh-Jabbari, M. M., \& Yin, L. (2021). Does Hubble tension signal a
breakdown in FLRW cosmology?. Classical and Quantum Gravity, 38(18), 184001.

\bibitem{D0} P. G. Fernandes, Charged black holes in AdS spaces in 4D
Einstein Gauss-Bonnet gravity, Phys. Lett. B, 135468 (2020).

\bibitem{D3} Fernandes, P. G., Carrilho, P., Clifton, T., \& Mulryne, D. J.
(2022). The 4D Einstein-Gauss-Bonnet theory of gravity: a review. Class.
Quant. Grav.

\bibitem{D8} M. Cveti\v{c}, S. I. Nojiri and S. D. Odintsov, Black hole
thermodynamics and negative entropy in de Sitter and anti-de Sitter
Einstein--Gauss--Bonnet gravity, Nucl. Phys. B, \textbf{628(1-2)}, 295-330,
(2002).

\bibitem{D7} P. G. Fernandes, Charged black holes in AdS spaces in 4D
Einstein Gauss-Bonnet gravity, Phys. Lett. B, 135468 (2020).

\bibitem{R9} D. Kastor, S. Ray and J. Traschen, Enthalpy and the Mechanics
of AdS Black Holes, Class. Quant. Grav., \textbf{26} 195011, (2009).

\bibitem{R10} D. Kubiz\v{n}\'{a}k, R. B. Mann and M. Teo, Black hole
chemistry: thermodynamics with Lambda, Class. Quant. Grav., \textbf{34(6)},
063001 (2017).

\bibitem{D4} N. Farhangkhah, New black hole solutions of Gauss-Bonnet
gravity in the presence of a perfect fluid, Iran. J. Phys. Res., \textbf{%
17(5)}, 729-735 (2019).

\bibitem{Z18} C. H. Wu, Y. P. Hu and H. Xu. Hawking evaporation of
Einstein--Gauss--Bonnet AdS black holes in D$\geq 4$ dimensions, Eur. Phys.
J. C, \textbf{81(4)}, 1-9 (2021).

\bibitem{MBOU2} Bousder, M. (2022). A new constant behind the rotational
velocity of galaxies. Journal of Cosmology and Astroparticle Physics,
2022(01), 015.

\bibitem{MBOU3} Bousder, M. (2021). Quantum f (R) gravity and AdS/CFT. arXiv
preprint arXiv:2104.14434.

\bibitem{MBOU4} Bousder, M., Sakhi, Z., \& Bennai, M. (2020). A new unified
model of dark matter and dark energy in 5-dimensional f (R, $\phi$) gravity.
International Journal of Geometric Methods in Modern Physics, 17(13),
2050183.

\bibitem{Z19} S. B. Giddings, Phys. Lett. B, \textbf{754}, 39-42 (2016).

\bibitem{PLB1} Barrow, J. D. (2020). The area of a rough black hole. Phys.
Lett. B, 808, 135643.

\bibitem{L1} Antoniou, G., Bakopoulos, A., \& Kanti, P., 2018, \textrm{Phys.
Rev. D}, 97(8), 084037.

\bibitem{L2} Maeda, K. I., Ohta, N., \& Sasagawa, Y. 2009, \textrm{Phys.
Rev. D}, 80(10), 104032.

\bibitem{L3} Fernandes, P. G. (2020), \textrm{Phys. Lett. B}, 135468.

\bibitem{L5} Doneva, D. D., \& Yazadjiev, S. S., 2021, \textrm{J. Cosmol.
Astropart. Phys.}, 2021(05), 024.

\bibitem{DM} Bean, R., Flanagan, E. E., \& Trodden, M. 2008, \textrm{Phys.
Rev. D}, 78(2), 023009.

\bibitem{L12} Khoury, J., \& Weltman, A., 2004, \textrm{Phys. Rev. D},
69(4), 044026.

\bibitem{L13} Folomeev, V., Aringazin, A., \& Dzhunushaliev, V., 2013,
\textrm{Phys. Rev. D}, 88(6), 063005.

\bibitem{L14} Matos, T., \& Urena-Lopez, L. A., 2000, \textrm{Class. Quantum
Gravity}, 17(13), L75.

\bibitem{L15} 4Arbey, A., \& Coupechoux, J. F., 2021, \textrm{J. Cosmol.
Astropart. Phys.}, 2021(01), 033.

\bibitem{L16} Aghanim, N., Akrami, Y., Ashdown, M., Aumont, J., Baccigalupi,
C., Ballardini, M., ... \& Roudier, G., 2020, \textrm{Astron. Astrophys.},
641, A6.

\bibitem{D9} M. Li, A model of holographic dark energy. Phys. Lett. B,
\textbf{603(1-2)}, 1-5 (2004).
\end{thebibliography}
\end{document}